\documentclass[envcountsect,envcountsame]{llncs}
\usepackage{amssymb,stmaryrd}
\usepackage{amsmath}
\usepackage{epsfig}

\title{\mbox{Dynamic Scope-Based Dijkstra's Algorithm}%
  \thanks{Research supported by the Czech Science Foundation,
    grants P202/11/0196 and Eurocores GIG/11/E023.}}

\author{Petr Hlin\v{e}n\'y ~\and~ Ondrej Mori\v{s}}

\institute{Faculty of Informatics, Masaryk University \\
  Botanick\'a 68a, 602 00 Brno, Czech Republic  \\
  \email{hlineny@fi.muni.cz, xmoris@fi.muni.cz}}

\begin{document}

\maketitle

\begin{abstract}
We briefly report on the current state of a~new dynamic algorithm 
for the route planning problem based on a concept of scope (the static 
variant presented at ESA'11, \cite{HM2011A}). We first motivate 
dynamization of the concept of scope admissibility, and then we briefly 
describe a modification of the scope-aware query algorithm of~\cite{HM2011A} 
to dynamic road networks. Finally, we outline our future work on this concept.
\end{abstract}

\section{Introduction}

The single pair shortest path problem in real-world road networks,
also known as route planning, has many important everyday applications.
There are two most significant variants of the problem -- static and dynamic
route planning.

In the \emph{static} variant, a road network is fixed during the computation
of optimal routes (\emph{query}).  Static route planning received a lot of
attention during the last decades.  On the other hand, in the \emph{dynamic}
variant a road network is subject to change in time -- some new roads are built,
some other are closed, traffic jams or car accidents happen, some
routes must be avoided due to turn angle limits, and so on.  Clearly, the
latter is a more realistic scenario.  Actually, even time-dependent cost functions
can be modeled in dynamic route planning to some extent.

Classical algorithms such as Dijkstra's \cite{Dijkstra1959} or A*
\cite{Hart1972} and their dynamic adaptions \cite{Cooke1966} are not well
suitable neither for static nor for dynamic variants because of huge road
networks size.  Thus, a feasible solution lies in computing suitable
auxiliary data of the road network (\emph{prepro\-cessing}) in order to
improve both time and space complexity of subsequent queries.  This
technique led to several very interesting static approaches in the last
decade, see extensive surveys, for instance, in
\cite{Schultes2008,Delling2009}.  Some of these algorithms such as
highway-hierarchies \cite{Sanders2006}, ALT \cite{Goldberg2005B} or, for
example, geometric containers \cite{Wagner2005} were proved to work fine
also in the dynamic scenario
\cite{Flinsenberg2004,Wagner2004,Delling2007,Schultes2007}.  

Recently, in order to fill a gap between a variety of exact route planning
approaches, we have published \cite{HM2011A} a different novel approach
aimed at ``reasonable'' routes.  It is based on a concept of scope, whose
core idea can be informally outlined as follows: 
The edges of a road network are associated with a {\em scope} map
such that an edge $e$ assigned scope $s_e$ is admissible on a route $R$ if,
before or after reaching $e$, such $R$ travels distance less than a value
associated with $s_e$ on edges with scope higher than~$s_e$.  The desired
effect is that low-level roads are fine near the start or target positions,
while only roads of the highest scope are admissible in the long middle
sections of distant routing queries. 
Overall, this nicely corresponds with human
thinking of intuitive routes, and allows for a very space-efficient
preprocessing, too.

\paragraph{New Contribution.}

In the dynamic scenario, however, a static scope map may badly fail.
Imagine, for instance, a closure of a motorway tunnel which can be bypassed
only on low-level mountain roads.
Then a detour would not be scope admissible in the aforementioned
sense, and so a dynamic adjustement of this definition is needed.
We present such an adjusted definition here,
along with a modification of Dijkstra's algorithm for scope admissible
routes in this dynamic scenario.
Our algorithm is exact, and its time complexity grows only slightly over
ordinary Dijkstra if few negative changes are introduced in the network.

% The core of aforementioned scope-based route planning approach is
% represented by a modification of Dijkstra's algorithm for scope admissible
% walks.  In this work, we adapt this algorithm for dynamic route planning
% scenario.  Our goal is to ensure that dynamic algorithm remains correct and
% exact in the sense that it finds provably optimal route w.r.t.  dynamic
% version of scope admissibility.  Such algorithm would have many interesting
% applications.

\section{Preliminaries}
\label{sec:prelimin}

A {\em directed graph} $G$ is a pair of a finite set $V(G)$ of vertices 
and a finite multi-set $E(G) \subseteq V(G) \times V(G)$ of edges. A 
\emph{walk} $P \subseteq G$ is an alternating sequence $(u_0,e_1,u_1,\ldots,e_k,u_k)$ 
of vertices and edges of $G$ such that $e_i = (u_{i-1},u_i)$ for $i = 1,\ldots,k$. 
The \emph{weight} of a walk $P \subseteq G$ w.r.t. a weighting $w: E(G) \mapsto 
\mathbb{R}$ of $G$ is defined as $|P|_w=w(e_1)+w(e_2)+\dots+w(e_k)$ where 
$P=(u_0,e_1,\ldots,e_k,u_k)$. An {\em optimal walk} between two vertices achieves 
the minimum weight over all walks.

A road network is referred to as a pair $(G,w)$ where $G$ is a directed 
graph (such that the junctions are represented by $V(G)$ and the roads by 
$E(G)$), and $w$ (cost function) is given as a {\em non-negative} edge 
weighting $w: E(G) \mapsto \mathbb{R}^+_0$. In the dynamic scenario, $w$ 
is simply replaced with $w^*$ (differing from $w$ only on few edges, say).
If $e$ is removed then let $w^*(e)=\infty$. 

Driven by real-world motivation, we focus on negative (increased weight,
even to $\infty$) changes in $w$. We thus now for simplicity omit the 
possibility of adding new edges to $G$, though we understand it may be 
useful when, e.g., a designated detour locally changes the road network.

\section{Scope and Scope Admissibility} 

A simplified version of the scope concept is briefly introduced here.
We strongly recommend reading \cite{HM2011A} for more detailed treatment
and, due to lack of space, omit most details here.

\begin{definition}[\cite{HM2011A}]
\label{def:scope}
Let $(G,w)$ be a road network. \emph{A scope mapping} is defined as ${\cal{S}}: 
E(G) \mapsto \mathbb{N}_0 \cup \{\infty\}$ such that $0,\infty \in 
Im({\cal{S}})$. Elements of the image $Im({\cal{S}})$ are called \emph{scope 
levels}. Each scope level $i\in Im({\cal{S}})$ is assigned a constant value of 
\emph{scope} $\nu^{\cal{S}}_i \in \mathbb{R}_0 \cup \{\infty\}$ such that 
$0 = \nu^{\cal{S}}_0 < \nu^{\cal{S}}_1 < \cdots < \nu^{\cal{S}}_\infty = \infty$.
\end{definition}

In practice there are only a few scope levels (say,~5). The desired effect, 
as formalized next, is in admitting low-level roads only near the start or 
target positions until higher level roads become widely available.

\begin{definition}[\cite{HM2011A}]
  \label{def:xadmissible}
  Let $(G,w)$ be a road network and $x \in V(G)$. An edge $e =   
  (u,v) \in E(G)$ is \emph{$x$-admissible} in $G$ for a scope mapping
  $\cal{S}$ if, and only if, there exists a walk $P \subseteq G - e$  
  from $x$ to $u$ such that
  \begin{enumerate}\parskip 3pt
    \item each edge of $P$ is $x$-admissible in $G - e$ for $\cal{S}$,
    \item $P$ is optimal subject to (1), and
    \item for $\ell={\cal{S}}(e)$,~
	$\sum_{f \in E(P),\, {\cal{S}}(f)>\ell}\, w(f) \,\le\> \nu^{\cal{S}}_{\ell}$.
  \end{enumerate}
\end{definition}
\begin{definition}[\cite{HM2011A}]
  \label{def:stadmissible}
  Let $(G,w)$ be a road network and $\cal{S}$ a scope mapping. For a
  walk $P=(s=u_0,e_1,\dots e_k,u_k)$ in $G$; $P$ is
  \emph{$s$-admissible} in $G$ for $\cal{S}$ if every $e_i \in E(P)$     
  is $s$-admissible in $G$ for $\cal{S}$.
\end{definition}

% Consider the following dynamic scenario -- there is an admissible walk and
% suddenly some if its edges increases its weight.  A driver is already on her
% way and reach such edge.  She should try to find a detour to bypass such
% problematic road and, still, such detour must minimize the costs and be
% reasonable.  Scope admissibility must be refined to capture such situations
% and some non-admissible edges near this changed edge should be allowed in a
% clever way.

\paragraph{Static $\cal{S}$-Dijkstra's Algorithm.}
\mbox{Aforementioned} seemingly complicated definitions can be smoothly integrated
into (the bidirectional variants of) Dijkstra's or A* algorithms, simply
keeping track of the extreme $s$-admissibility (or $t$-adm.\ in reverse dir.) condition:
\begin{itemize}\parskip 3pt
\item
  For every accessed vertex $v$ and each scope level $\ell\in Im({\cal{S}})$,
  the algorithm keeps, as $\sigma_\ell[v]$, the best achieved value 
  of the sum $\sum_{f \in E(P),\, {\cal{S}}(f)>\ell}\, w(f)$.
\item
The $s$-admissibility of edges $e$ starting in $v$ then depends
  just on $\sigma_{{\cal S}(e)}[v]\leq \nu^{\cal{S}}_{{\cal S}(e)}$, and only 
  \mbox{$s$-admissible} edges are relaxed further.
\end{itemize}

\begin{theorem}[\cite{HM2011A}]
\label{thm:SDijkstra}
{\em$\cal{S}$-Dijkstra's algorithm} (uni-di\-rectional), for a road 
network $(G,w)$, a scope mapping ${\cal{S}}$, and a start vertex 
$s \in V(G)$, computes an optimal $s$-admissible walk from $s$ to 
every $v \in V(G)$ in time ${\cal O}\big(|E(G)|\cdot|Im({\cal{S}})|+
|V(G)|\cdot\log |V(G)|\big)$.
\end{theorem}

The most important computational aspect of scope lies in the fact that only
the edges of {\em unbounded} scope level $\infty$ matter for global preprocessing
(an idea related to better known {\em reach} \cite{Gutman2004}).
Informally, the query algorithm of \cite{HM2011A} works in stages:
In the \emph{opening cellular phase}, the road
network is locally searched (uni-directional $\cal{S}$-Dijkstra) 
from both start and target vertices
until only edges of unbounded scope are admissible.
Then a small preprocessed ``boundary graph'' is searched by another algorithm
(e.g.\ hub-based labeling \cite{Abraham2011}) in the \emph{boundary phase}. 
Finally, in the \emph{closing cellular phase}, the scope-unbounded long 
middle section of the route is ``unrolled'' in the whole network.

We remark that the boundary graph will remain static even in the dynamic 
scenario (due to expensive preprocessing), and dynamic changes will be
mainly dealt with in the closing cellular phase. Yet we have to pay 
attention to scope admissibility since it is the key to much improved 
preprocessing \cite{HM2011A} to the boundary graph. It is therefore 
essential to ``dynamize'' our definition and $\cal{S}$-Dijkstra's 
algorithm for that purpose.

\section{$\cal{S}$-Dijkstra's Algorithm -- Dynamization}

In the rest, due to limited space, we only briefly sketch the uni-directional
{\em Dynamic $\cal{S}$-Dijkstra's Algorithm} used locally in the opening cellular phase
(while the admissibility definition is implicitly embedded in it).
This procedure can be routinely turned into bidirectional and then
used to resolve dynamic changes in cells during the closing cellular phase.

We first remark on the ``only negative change'' assumption of our approach
(Sec.~\ref{sec:prelimin}).
This well corresponds with a real-world situation in which just ``bad things
happen on the road'', and the driver thus usually has to find an available
detour, instead of looking for unlikely road improvements.
Therefore, we are content if our query algorithm finds that an optimal route of
the original network (wrt.~$w$) is admissible, though not perfectly optimal,%
\footnote{Note that a designated detour of a road construction may perhaps 
turn out faster than another previously optimal route.}
in the changed network ($w^*$).
However, when things go worse with $w^*$, 
then our algorithm will always find an optimal
dynamic-scope admissible detour in the changed network.

\paragraph{Main Informal Idea.} 

Imagine a driver approaching a road restriction or closure. What would she do?
% A~driver reached an unexpected change\footnote{The weight $w(e)$ of
% $e=(u,v)$ increased its weight to $w(e)^*$.}.  What is her thinking in such
% a situation?
Intuitively, the best solution is for her to slip off the original route
(even ahead of the restriction), and re-allow the use of low-level (i.e.,
inadmissible in the ordinary setting) roads nearby the restriction.
% She does not want to return back to the start $s$ too far. If possible, she
% would like to continue her $s$-admissible comfort route from $u$. 
% Otherwise, she is ready even to slip to worse local non-$s$-admissible
% roads.  She wants to return to her normal predictable route (when $w=w^*$)
% as soon as possible.  
Of course, she still wants to minimize detour costs and drive reasonably in 
terms of such adjusted scope admissibility.

\paragraph{Triple Search.} 

Dynamic changes in our $\cal{S}$-Dijkstra's algorithm, starting from $s$,
are resolved by a \emph{detour procedure} executed whenever an $s$-admissible
changed edge $e=(u,v)$, i.e.\ with $w(e)<w^*(e)$, is going to be relaxed.
For simplification we assume that there is only one such changed edge~$e$ in the
whole network.

The detour procedure is analogical to ordinary Dijkstra, except that to its 
single (called \emph{live}) search %queue 
it adds two other auxiliary searches %queues, 
called \emph{dead} and \emph{detour}. Their roles are as follows:
% shortly summarized here:

% Before reaching $e$, \mbox{$\cal{S}$-Dijkstra's} algorithm maintains a
% single search and its queue $Q$ for $s$-admissible reached vertices:
% When $e=(u,v)$ is relaxed and its end vertex $v$ is updated, $v$ is removed
% from $Q_{live}$ and two new similar searches are started as follows:
\begin{itemize}\parskip 3pt

\item \textit{Live} (the original% queue
) -- 
running as in static \mbox{$\cal{S}$-Dijkstra}, relaxes only $s$-admissible 
edges while using dynamic $w^*$. Let $Q_{live}$ denote its queue of 
discovered vertices, $d_{live}$ its temporary distance estimates, and 
$\sigma_{live}$ its scope condition vector.

\item \textit{Dead} -- 
% represents original search producing routes as if no edge changed, it is
% similar to previous search, the only difference is that it uses the original
% (but no longer actual -- dead) weighting $w$.  
started from the end $v$ of $e$ as if $w(e)$ was not changed.
So, initially, $Q_{dead}= \{v\}$ and $d_{dead}=d_{live}$ except
$d_{dead}[v]=d_{live}[u]+w(e)$.
The purpose of $Q_{dead}$ is to later identify which alternative walks
are actual detours for~$e$.

\item \textit{Detour} - 
the core new search started from $u$.
Upon reaching $e$, it resets $\sigma_{detour}[u]$ to $0$
on all scope levels $\leq{\cal{S}}(e)$.
%% timto si nejsem zcela jisty, mozna resetovat vsechny scope levels %%
Then it fills $Q_{detour}$ with $u$
and vertices $x$ on the access route from $s$ to~$u$ such that
a reverse search from $u$ to $x$ does not exhaust $\sigma_{detour}$ yet.
Note that $\sigma_{detour}$ is normally updated with this reverse search.
However, $d_{detour}[x]=d_{live}[x]$ for those added $x\in Q_{detour}$.
%  it relaxes only
% \mbox{$u$-admissible} edges, uses the dynamic weighting $w^*$.  It has one
% specific property -- \emph{debit} (sketched later).  It starts with
% $Q_{detour} = \{u\}$, $d_{detour}[u]=d_{live}[u]$, $d_{detour}[v]=\infty$
% for all $v \in V(G) \setminus \{u\}$, $\pi_{detour}[u]=\pi_{live}[u]$ and
% $\sigma_{detour}[u]=(0,\ldots,0)$.

\end{itemize}

% \paragraph{Step.} 
The search then continues concurrently with all the three queues
(so, starting turns will likely be taken by $Q_{detour}$).
Every relaxation from one of the queues is done as in the static
$\cal{S}$-Dijkstra's algorithm, i.e., 
updating also the appropriate $\sigma_\bullet$ vector.
Rules which relate together the three searches are outlined below.

\paragraph{Live or Dead.} 

% In case of tie, dead search goes first, detour next and live search is the last.
Our driver's desire is to get back to her original route represented by the dead
search.  This happens when the dead search meets either with the live search
(no need to bypass the problematic edge~$e$) or with the detour search
(a detour is found).

\begin{figure}[tb]
  \centering
  \centerline{\epsfig{file=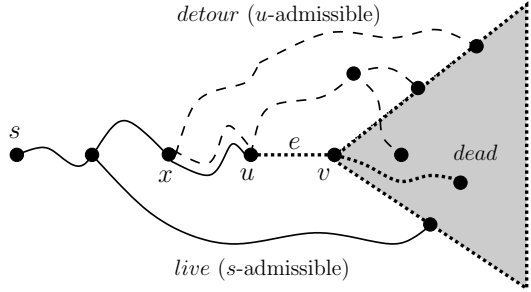, scale=0.7}}
  \caption{\footnotesize
An illustration of our detour procedure. 
% $\cal{S}$-Dijkstra's algorithm relaxed $e$ and decreased $d[v]$ to $d[u]+w(e)$.
% Then the weighting of $e$ increased from $w(e)$ to $w^*(e)$.  
Upon reaching dynamically changed $e$ such that $w^*(e)>w(e)$;
the original live search continues wrt.\ $w^*$ from already reached vertices
except~$v$ (solid lines), and two new searches are started --
the detour search from $u$ and some of its predecesors (in dashed lines)
and the dead search continueing after $v$ wrt.\ $w$ (dotted area).}
  \label{fig:detour}
\end{figure}

For more details (see also Fig.~\ref{fig:detour}),
imagine a vertex $y\in V(G)$ being relaxed from one of the three queues.
\begin{itemize}\parskip 3pt
\item
Suppose $y$ is relaxed from $Q_{live}$.
If $d_{dead}[y]\geq d_{live}[y]$ or $d_{detour}[y]\geq d_{live}[y]$,
then $y$ is removed from $Q_{dead}$ or $Q_{detour}$, respectively.
\item
Suppose $y$ is relaxed from $Q_{detour}$.
If $d_{dead}[y]\leq d_{detour}[y]< d_{live}[y]$, then 
$y$ is moved from $Q_{detour}$ (implic.\ with all its descendants) 
into $Q_{live}$ setting new distance estimate $d_{live}[y]:=d_{detour}[y]$.
\item
Suppose $y$ is relaxed from $Q_{dead}$.
If $d_{detour}[y]\leq d_{dead}[y]< d_{live}[y]$, then again, 
$y$ is moved into $Q_{live}$ with new distance estimate
$d_{detour}[y]$.
\end{itemize}
Notice that whenever $Q_{dead}$ or $Q_{detour}$ becomes empty,
the other one may also be removed and the algorithm then continues as
original $\cal{S}$-Dijkstra.

\renewcommand{\figurename}{\footnotesize Fig.}

\smallskip
Altogether, the above described Dynamic $\cal{S}$-Dijk\-stra's Algorithm
adds at most a constant multiplicative factor to the complexity of static
$\cal{S}$-Dijkstra, and we propose that usually this increase is only by 
an additive factor (the dead and detour searches restricted to a neighbourhood of $e$).

% A vertex might has a state -- it can be either live or dead, initially 
% $u$ is live, $v$ is dead and all $w \in Q_{live}$ are live, the other vertices have no state. When a~vertex $w \in V(G)$ is reached and:
% \begin{itemize}
% \item \emph{$w$ has no state} -- it inherits the state of its parent;
% \item \emph{$w$ is dead} -- if $w$ is now reached by dead search, then $w$ become dead, otherwise $w$ is now reached either in live or detour search and $w$ becomes live, moreover if $w$ is now reached by detour search, then $d_{live}[w]$ is set to $d_{dead}[w]$;
% \item \emph{$w$ is live} -- if $w$ is now reached in live of detour search, then $w$ remains live, otherwise $w$ is now reached in dead search and $w$ becomes.
% \end{itemize}
% In dead search, if a vertex $w$ is live, its outgoing edges are not relaxed and it becomes scanned (in dead search) immediately. In other words, dead search terminates in $w$. It is because live $w$ can be already reached by detour or by live search.
% \paragraph{Termination.}
% When $Q_{dead}$ and $Q_{detour}$ are empty, then our detour procedure is over and main $\cal{S}$-Dijkstra's algorithm continues as usually with $Q_{live}, d_{live}, \pi_{live}$ and $\sigma_{live}$. 

\paragraph{Borrowing Scope in Detour.} 
There is one more specific aspect of the aforementioned detour search.
We not only want to reset $\sigma_{detour}[u]$ upon reaching changed $e$
in the forward direction, but we intend to do the same for $\sigma_{detour}[v]$ ``backwards''.
Informally, we would like to allow low-level roads not only to slip off
the original route, but also to return to it from a detour.
However, this cannot be done simply in a backward search,
and so we instead ``borrow'' a needed scope value for $\sigma_{detour}$.

Precisely, the detour search is allowed to relax even non-$s$-admissible
edges, keeping track of the limited scope value {\em debt} (on each level).
This debt must then be repaid ``from $v$'' when the detour search meets 
the dead search (if it is not repaid in full, then this search branch
subsequently dies).
Again, due to lack of space, we omit further details.

%  we allow it to use non-$u$-admissible if detour was not yet found and there
% is no more $u$-admissible edges.  The distance driven through each such
% non-$u$-admissible walk is called \emph{debit} and it must be compensated
% when a detour is found.  Informally, suppose that detour and dead search
% meet at vertex $w$ and it is reached with debit $d$ in detour search.  Then
% the debit $d$ of $w$ is \emph{compensated} if $d_{dead}[w] - d_{dead}[v] \ge d$.  

\paragraph{Multiple Changes.} 

The previous dynamic algorithm may be extended to handle multiple changed
edges in $w^*$, too, as we very briefly outline now.
We introduce multiple dead and detour searches, each labeled by a set of all
changed edges that affected it.
In this view, the original live search is actually the dead search with
the empty label.

All these concurrent searches are related together by a complex set of rules
depending on their label sets
(such as, finishing an $L$-labeled detour of an edge $e_1$ moves it to the
search labeled by $L\setminus\{e_1\}$; etc).
We summarize:

\begin{theorem}
\label{thm:DSDijkstra}
{\em Dynamic $\cal{S}$-Dijkstra's algorithm} (uni-directional), for a road 
network $(G,w)$ dynamically changed to $(G,w^*)$, a scope mapping ${\cal{S}}$, 
and a start vertex $s \in V(G)$, computes a dynamically $s$-admissible walk from $s$ to 
every $v \in V(G)$.
This computed walk is optimal in $(G,w^*)$, or in $(G,w)$.%
\footnote{This slick formulation is to handle the (unlikely in practice)
situation when the changed network actually contains a shorter route from
$s$ to $v$ which may not be found.}

If $c$ denotes the number of edges $e$ such that $w(e)<w^*(e)$,
then the algorithm runs in time at most
${\cal O}\big(2^c\cdot(|E(G)|\cdot|Im({\cal{S}})|+
|V(G)|\cdot\log |V(G)|)\big)$.
\end{theorem}
Even though the factor $2^c$ may look horrible, we believe
the actual effect on time complexity is marginal in real-world scenarios
with not-so-many dynamic changes (due to typical ``locality'' of detours).
% (see the above considerations).
A thorough experimental evaluation of the complexity of our algorithm is the
subject of ongoing research.

% It might happen that in our detour procedure any of three searches reaches another changed edge. Obviously, we can safely ignore any number of changes in dead search due to its purpose, only debit compensation is adjusted. In case of live and detour search, these changes must be handled in a clever way. Nested detour procedures are not allowed. Due to lack of space, we omit further details here and just simply claim that such ``clever way'' does exist. 
% \paragraph{Bidirectional Version.} It is very easy to implement aforementioned dynamic $\cal{S}$-Dijkstra's algorithm bidirectionally. As usually, two dynamic $\cal{S}$-Dijkstra's algorithms are executed at the same time -- first in forward direction from the start $s$ and another from the target $t$ in reverse direction (i.e. in reversed road network). Both searches use aforementioned detour procedure. There can be only one such procedure running at the same time and it ignores search in another direction.

\section{Discussion}

We have outlined the current state of our work on dynamization of the
scope-base route planning technique \cite{HM2011A} for both unexpected and
predictable (to some extend) road network changes.  Our approach is aimed at
a proper relaxation of scope admissibility when a driver approaches changed road
segment, by locally re-allowing nearby roads of lower scope level.  At the
same time we claim that the computed detour minimizes costs and still remains
reasonable in terms of scope admissibility.  However, formalized algorithm,
its complexity analysis, rigorous proof of correctness and most details are
omitted due to lack of space.

In a summary, we have shown that a scope-based route planning approach 
with cellular preprocessing \cite{HM2011A}
can be used not only in static but also in dynamic road networks.
Our immediate future work in this direction will include the following points;
\begin{itemize}\parskip 3pt
\item
precise definition of dynamic scope admissibility,
\item
adding new edges and positive dynamic changes,
\item
incorporating so called maneuvers, and
\item
experimentally evaluating this dynamic algorithm
on real-world map data.
\end{itemize}

% In this paper we sketched how to modify the uni-directional scope based
% $\cal{S}$-Dijkstra's algorithm for the most important dynamic scenarios --
% edge weights increases or even edge removals (i.e.  vertex removals are
% covered too).  Then we briefly mentioned how to use such modified
% algorithm bidirectionally.  Thus we have actually shown that scope-based
% route planning approach can be used not only in static but also in dynamic
% road networks.

% Welcome added benefits of our approach is that it allows to re-route already prepared easily in case of changed weight. The next steps of our research will naturally be:
% 
% \begin{itemize}\parskip-2pt
% \item Refine the definition of scope admissibility to capture necessary detours by rigorous mathematical language.
% 
% \item Handle positive changes such as edge additions of edge weights decreases in order to avoid suboptimal route planning solutions.
% 
% \item Incorporate various local restrictions and traffic regulations modeled by so-called \emph{maneuvers} \cite{HM2011B}, this is a quite hard but unexplored and very promising area.
% 
% \item Extend the dynamic road network for the time-dependent scenario, where a cost function depends on time everywhere.
% 
% \item Finally, the most important work to be done is to experimentally evaluate proposed algorithms on real-world road networks and their comparison with other dynamic route planning approaches.
% \end{itemize}

\small 
\bibliographystyle{abbrv}

\end{document}